# Hydrodynamic Lift on Boats

Joseph L. McCauley
Physics Dept.
Univ. of Houston
Houston, Tx. 77204
jmccauley@uh.edu

## Abstract

Dimensional analysis is a very powerful tool in hydrodynamics. It's more general and more useful than scaling, which is an approximation limited to a small range of other often hidden dimensionless parameter values. By using the lift coefficient and the Froude number, both dimensionless, we offer clear definitions of lift onset and planing. Boats float supported only by buoyancy at rest and at very low speeds. As the speed increases, the bottom of a boat capable of planing makes a sharp transition to a lifting surface while plowing up the bow wave. As the speed increases further the boat planes, and at a high enough speed buoyancy becomes negligible; hydrodynamic lift carries the weight. More generally, any flow past a lifting surface generates a force perpendicular to the flow once the flow separates from the trailing edge without backflow (without vortex formation). Wings, hydrofoils, sails, and propellers are examples of lifting surfaces. The idea of a planing hull as lifting surface has been discussed inconclusively and unconvincingly in the literature. We provide visual evidence that the onset of lift is a sharp transition that occurs before planing, and also offer an empirically correct lift coefficient for the planing of a v-bottom hull (prismatic shape) at high enough Froude number that the wet surface area is confined roughly to a triangular shape near the transom. Our lift coefficient is suggested by our calculation of the lift on a fully submerged delta wing hydrofoil. The lift is reduced once the hydrofoil pierces the air-water interface.

## 1. Introduction

We discuss flows over boat bottoms, so some boat terminology is necessary. Hull means bottom and sides of a boat. The hull is partly or fully covered on top by the deck. Our interest is in the hull's exterior shape beneath the waterline, and the waterline varies with the speed of a planing hull. The bow of a boat is the front while the stern is the rear. Aft means toward the rear, fore means toward the front. Trim angle $\alpha$ means the angle between the water's surface and the boat's bottom. The beam is the width of the boat. Most important is on the geometry of the boat near the stern, in particular the shape of the wet section of hull at the stern. Ships, efficient rowboats, canoes, and slow sailboats may be more or less symmetric fore and aft with a sharp line as both stern and bow as in fig. 1. In these cases the beam at the stern is zero. Such a hull has a streamline form designed to prevent the separation of the flow from the sides and bottom of the boat at low enough Reynolds nrs. [1]. Separation of the flow from the hull would cause eddy/vortex formation and therefore increased form drag, where net drag is the form drag plus skin friction [1]

$$\vec{F} = \iint_S P\hat{n}\,dA + \iint_S \vec{\sigma}\cdot d\vec{A} \tag{1}$$

with

$$\sigma_{ik} = \mu\left(\frac{\partial v_i}{\partial x_k} + \frac{\partial v_k}{\partial x_i}\right). \tag{2}$$

The net force on an object in a flow field is $\vec{F}$, the fluid velocity is $\vec{v}$, P is the pressure and $\vec{\sigma}$ is the stress tensor (2). The first term includes the form drag (antiparallel to the flow) and the lift (perpendicular to the flow). The second term in (1) is the skin friction. The thin turbulent boundary layer doesn't contribute to understanding the physics of lift on wings or planing hulls once the required vortices have formed [2]. The most mathematically difficult aspect of planing would be to calculate wave and spray formation at the air-water interface. A crude attempt to approximate this imagines the waves and spray resulting from dropping the boat vertically into the water [3]. For boats designed to plow through the water the streamline shape of the hull minimizes the form drag at low speeds where the trim angle remains zero. In this case the lift vanishes because lift on an uncambered surface requires a finite trim angle, or angle of attack α [2], and a planing hull bottom should be straight, camber-free, on the wet running surface. The bow wave is small and also represents form drag.

A boat with no beam at the stern (fig. 1) cannot plane and is thereby limited to very low speeds. As the speed is increased the size of the bow wave increases as the boat tries to ride up onto it causing the trim angle to become positive. A positive trim angle cannot generate planing if the beam of the bottom at the stern is of zero or too small width. We focus on boats that are capable of planing, capable of climbing onto and over the bow wave to glide over the water's surface and achieve much higher speeds than would be possible when buoyancy carries the weight. After planing the bow wave moves aft and the wet surface area is much reduced, reducing the drag. Skin friction is less that 10-15% of the total drag if boat bottom is smooth and straight [4] and the boat rides on a small area near the transom. Form drag (including waves and spray at the interface) dominates the total drag. Buoyancy carries less and less weight as the speed increases above planing speed. One might bring a canoe on top of the water (very unstably) by pulling it with a motorboat, but a boat like that in fig. 1 cannot rise out of the water and glide under any reasonable circumstances. A boat of the same size and weight with the right hull design, however, can easily plane with adequate power. A properly designed planing hull with the weight of a canoe would be fast, 30 mph or more, even with a small outboard motor of 5-7.5 hp.

In order to plane a hull must have a transom, meaning a finite beam of running surface at the stern with a sharp delineation of bottom and transom (fig. 2). Planing hulls are very inefficient when plowing through the water because of backflow up the transom resulting from eddy formation where the bottom and sides meet the transom. You can easily see the backflow if you paddle a planing hull in a region where dead sea grass floats in the water. As you paddle, the sea grass will follow the boat at the transom, you will have trouble getting rid of it. The backflow is eddy creation due to flow separation [1] at the bottom and sides. The form drag is much-increased by the backflow at the transom at very low speeds. But the sharp trailing edge made by the transom at the bottom along the finite beam is the feature that allows the boat to rise on top of the bow wave and then glide over the surface of the water as the speed is increased. Climbing onto the bow wave and planing is necessary in order to attain high speeds.

Camber is desirable on a hydrofoil but not on a planing hull bottom. The wet planing area ('running surface') of a planing hull should be straight, not rockered (convex) or hooked (concave). A planing hull should only be rockered well fore of the running surface. Sailboats and rowboats, on the other hand, may benefit from bottoms rockered toward the stern.

Sailboats with the appearance of a transom generally have a strongly rockered bottom so that the transom is always above the waterline. A hull built for high speeds (fig. 2) will run on very little wet surface area near the transom, on the order of a square foot or a few square feet. The 'bow wave' has moved very far aft in this case and the center of pressure is well aft of the center of gravity.

Here's our starting point for discussing lift vs. buoyancy and the onset of lift. As a boat like that of fig. 1 slowly moves forward you may observe the flow past the stern. There is a small bow wave but the water follows the contour of the boat and creates very little wake at the stern. In particular, there is no separation and backflow along the sides of the boat so long as the speed is low. In contrast, if you look over the transom of a planing hull as it moves forward at idle speed then there's a big backflow: the water follows the transom, the water is dragged

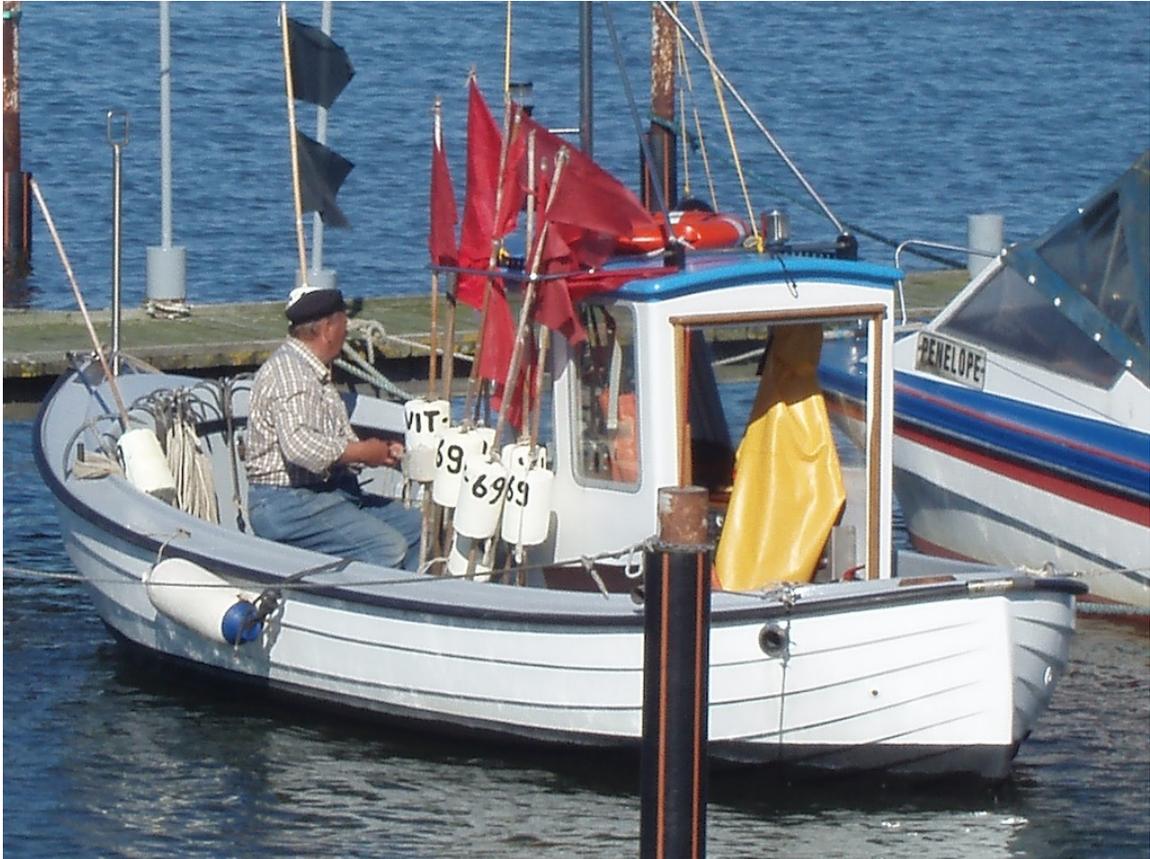

Fig. 1 shows a typical boat made for low-speed plowing. The stern has no transom (the beam vanishes at the stern) and the bottom is rockered.

by the boat (fig. 3). The trough behind the transom that would have been created by the boat's forward motion is continually filled by an eddy at the trailing edge of the bottom and sides. This is the analog of the starting vortex in Prandtl's flow visualization of a flow past a wing [5]. When lift develops then you see the trough, the entire transom is dry all the way to the bottom (fig. 4) and the water leaves the bottom as a sheet (a vortex sheet).

Hydrodynamics teaches us that any flow past a sharp edge creates a vortex, an eddy. Creation of a vortex along the surface or at the trailing edge of an object increases the form drag. Vortex creation occurs at the lowest Reynolds nr. in a flow past a sharp edge, like the trailing edge of a wing or the edge where a boat bottom meets the transom. There, the flow takes the form of a backflow from bottom to top of the wing, or up the transom from the bottom and sides of a boat. For a wing, as the speed increases the trailing edge vortex breaks free and goes downstream, at which point the flow separates cleanly from the wing's sharp trailing edge. This causes a net lift due to the vortex bound to the wing [2,5]. In both two- and three-dimensional streamline theory of flow past a wing one must impose the Kutta the condition mathematically to get rid of the backflow and generate lift [1,2]. This is an approximation demanded by streamline theory where vortex formation in the thin, turbulent boundary layer is neglected. In a real three-dimensional flow the sharp, eddy-free separation of the flow from the wing's trailing edge occurs naturally as the airplane's speed increases. We provide visual evidence in fig. 3 and 4 that this is also true for the vortex that causes the backflow at the transom of a boat. As the speed increases the trailing edge vortex cannot remain attached to the wing because it is not bound, it responds to the lift force on it due to the free stream flowing past and is pulled downward from the wing then carried away by the free stream [4]. An analogous one-time shedding of the trailing vortex occurs at lift onset on a planing hull. That the Kutta condition is in effect is reflected by the fact that only *one* vortex is formed and shed, more vortices are not continually reformed and shed. This is the empirical evidence for the Kutta condition.

## 2. Lift and the Kutta Condition

Early in the 20th century L. Prandtl developed boundary layer theory [1]. He created another tour de force by explaining the lift on a wing via circulation and shedding the starting/trailing vortex [5,6]. He produced a beautiful photograph taken in the fluid reference frame (reproduced in [5]) that shows the bound vortex forming at the leading edge of the wing along with the trailing vortex of opposite sign, the backflow from bottom to top of wing at the trailing edge. When the trailing vortex slides off the fuselage and becomes free [4] then it's carried downstream in the free stream, and the wing experiences a net lift due to the vortex bound to the wing (the wing's surface is the vortex core). The trailing and leading edge vortices develop at the same time because they are in fact the same vortex line wrapped into a horseshoe shape with both ends ending at different places on the fuselage [4]. The reason for this is circulation conservation: a vortex cannot end in the fluid, it must end on a boundary of the fluid [1]. The formal mathematical condition for no backflow at the trailing edge is called the Kutta condition [1,2]. A wing or propeller blade can be decomposed into a thickness function and the mean camber surface [2]. The mean camber surface generates all the lift, the thickness contributes nothing. The main point expressed by the Kutta condition mathematically is that the mean camber surface of a thin wing at a small angle of attack becomes a bound vortex sheet [2], with the sheet continuing off the wing as a free vortex sheet wake by circulation conservation. The free vortex sheet is unstable and rolls up to become the tip vortex seen in flow visualization pictures that are common on the internet. The main drag on the wing is the induced drag and is calculated from the tip vortex [1,2,4]. A planing hull bottom becomes a vortex sheet (velocity discontinuity) once the backflow up the transom ends, but the main drag is not from the analog of the tip vortex. The important drag comes from the lateral fluid motion represented by the side spray and waves at the air-water interface. This is at least an order of magnitude harder to calculate than is the induced drag for a fully submerged hydrofoil. These factors make the calculation of lift and drag in planing a much more difficult problem mathematically than the problem of a fully submerged hydrofoil, as I illustrate below.

Nearly every elementary physics text superficially explains the origin of the lift on a wing as merely due to a pressure difference over the top and bottom of a wing. The people who write such texts have inadequate knowledge of hydrodynamics. But without a net circulation about the wing the pressure difference integrated over the bottom and top of the wing is zero. This is an example of D'Alembert's Theorem [2]. The net circulation $\Gamma$ about the wing provides the pressure difference required for lift. To a zeroth approximation, a wing with lift is simply a bound vortex with circulation Γ [2]. In any case the net lift on the vortex bound to the wing is $F_L = \rho \Gamma U$ where Γ is the circulation, U is the air speed far from the wing, and ρ is the fluid density [2].

## 3. Lift and planing

A wing, hydrofoil, or any fully submerged lifting surface has a streamlined leading edge and a sharp trailing edge (a surface piercing propeller (fig. 2) has a sharp leading edge and generally has a thicker trailing edge). The sharp trailing edge of a planing hull is the line where the bottom meets the transom, and for this reason the bottom and transom should be separated by a sharp, well-defined angle, not by a rounded edge. Boat racers and other high performance boaters are often aware of this fact even if they have no idea of wing theory. Just as the trailing edge of a wing plays the essential role in lift-creation for flight, a finite beam of running surface at the stern (the transom) is a necessary condition for lift and planing. But whereas camber (curvature) increases the lift on a wing or the thrust generated by a propeller blade, a planing hull should have a wet running surface near the transom that is perfectly straight along the direction of motion. Otherwise the boat will misbehave and lose speed. A rocker (convexity) will cause porpoising, while a hook (concavity) will force the bow into the water. Both increase the wet running surface area and the drag.

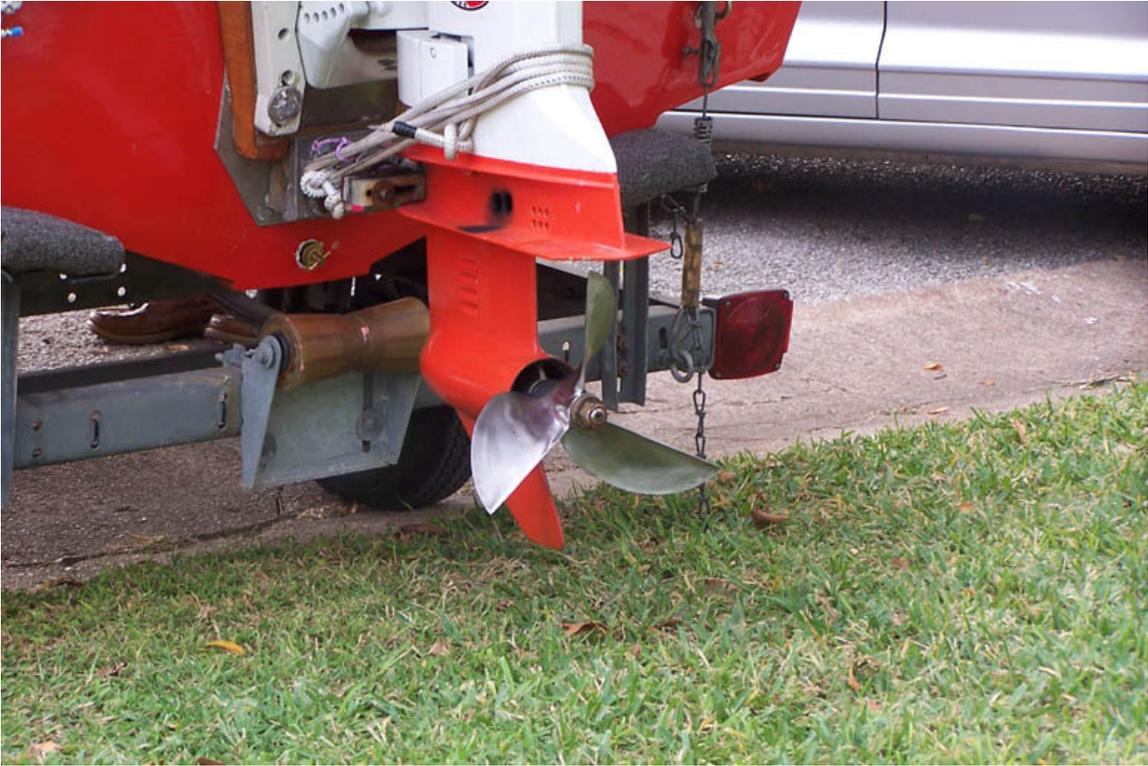

*Fig.* 2 shows the sharp trailing edge of a 14' 'pad v-bottom' Allison R14 boat at the transom. This boat was designed for high speeds, nearly 70 mph with 75 hp.

The location of the center of pressure is very important in the theory of lifting surfaces. The center of pressure is the ratio of the lift-induced torque to the lift force [2]. The net lift can be treated as if it acts at the center of pressure, analogous to treating gravity as if it acts at the center of mass. The exact location of the center of pressure depends on camber (curvature along the flow direction), but unless camber dominates the angle of attack then the center of pressure lies approximately 1/4 chord behind the leading edge of a fully submerged lifting surface like a wing or hydrofoil [2]. This is the basis for designing stable rudders and weathervanes, lifting surfaces that don't tend to 'flop' 180° when tilted at a finite angle of attack relative to the incoming flow. The location of the leading edge of the flow on a wing is clear, and on a boat bottom can be replaced approximately by the location of the dividing streamline, the streamline that ends fore on the bottom with zero speed. Clement [8], whose work agrees with our lift coefficient below, obtained empirical evidence that the center of pressure is about .77 to .88 the wet length l of a boat bottom for aspect ratios ranging from .5 to 5. The aspect ratio is $Æ = b^2 / A$ where A is the wet surface area and b is the wet length of beam at the transom. In wing theory the larger the aspect ratio the more 2-dimensional is the flow and the less is the induced drag. The largest possible lift coefficient is for two dimensional flow, $c_L = 2\pi\alpha$ . Achieving this with a boat bottom would produce maximum lift, and this can be accomplished with parallel keels on a flat bottom boat or (for airlift) the walls of the sponsons on a tunnel boat. I have run a small boat with nearly parallel off-center keels in a Baltic bay where dead sea grass is plentiful. The 330 lb boat (dry weight, empty and unrigged) ran 26.5 mph with a 1981 Evinrude 15 hp outboard, and the parallel keels directed the seagrass directly into the propeller. A later boat with a v-bottom had no such problem, the lateral flow off the bottom washed the seagrass to the side and never into the propeller.

Dimensional analysis tells us that the lift force has the form

$$F_L = \frac{\rho}{2} c_L A U^2 \qquad (3)$$

where $\rho$ is fluid density, A is the (wet) lifting surface area, and the lift coefficient $c_L$ is dimensionless. The lift coefficient need not be constant but may depend on dimensionless scaling parameters like the Froude nr. defined below. This formula agrees with the lift per unit wingspan written as $F_L = \rho \Gamma U$ because the integral for the circulation $\Gamma$ scales like U. The physics of the flow is hidden in the lift coefficient. To a zeroth approximation, the bottom of a boat is like the bottom half of a wing, the flow over the bottom is a vortex sheet with half the circulation of a submerged hydrofoil of the same shape. Now for the details of lift onset and planing.

We know from wing theory that $c_L \propto \alpha$ [2] where in our case $\alpha$ is the trim angle of the boat bottom (the running surface near the transom) relative to the water. When the boat slowly plows at idle speed then $\alpha=0$ and $F_L = 0$; the entire weight of the boat is carried by buoyancy. It is very instructive to lean over the transom and watch the change in the flow at the transom as the boat accelerates very slowly and incrementally from idle speed. Eventually, at a slightly greater speed than idle speed the boat tries to climb up the bow wave creating $\alpha>0$. One condition for lift is therefore satisfied, but lift has not yet developed: the entire transom is still wet (the Kutta condition is not satisfied). Fig. 3 shows the transom at a speed U slightly less than the speed where lift develops. Lift develops at a critical speed where suddenly the transom becomes completely dry as in fig. 4 where

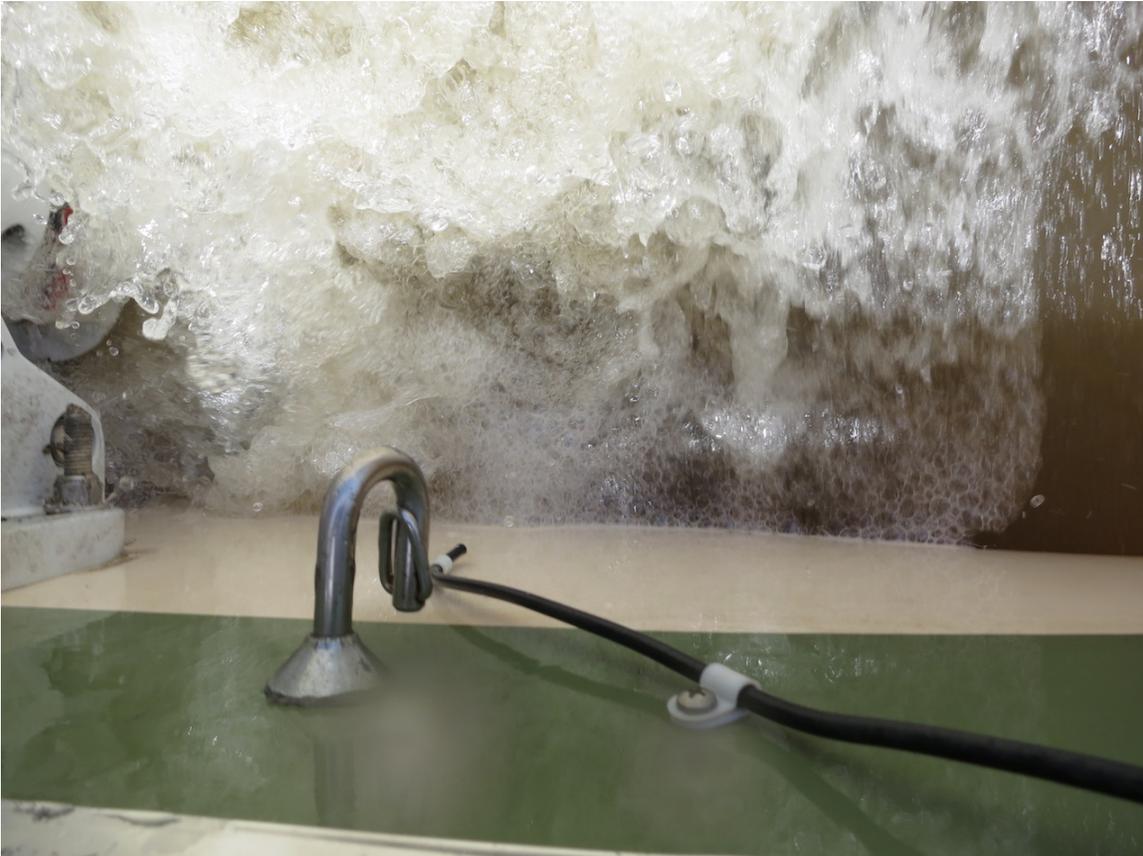

*Fig. 3* shows the backflow up the transom as the boat plows at a speed a few mph below that for the onset of lift. The backflow is due to the trailing edge vortex.

the water separates cleanly from the boat bottom at the transom. The backflow has disappeared and the water takes the form of a sheet leaving the bottom. This is analogous to the free vortex sheet continuing off the bound vortex sheet of a wing where the Kutta condition is satisfied. Figures 4 and 5 were both taken at the speed where lift suddenly begins. At a still greater speed the boat climbs on top of the water (fig. 6) and planes/glides rather than plowing. I define planing speed experimentally as the slowest speed where the boat glides over the water before a very slight reduction in throttle allows the boat to sink into the water at the transom and plow. For fast hull designs, at a speed greater than planing speed, the trim angle falls to a smaller angle that remains fixed as the speed is increased further, as in fig. 7. This last regime, I call 'clean planing': the trim angle is reduced but the boat rides higher on the water. At this speed buoyancy carries less than 15% of the weight. As the throttle is increased the boat rides higher above the water and gains speed, but at the same trim angle (as long as air lift is negligible). We can quantify this sequence as follows.

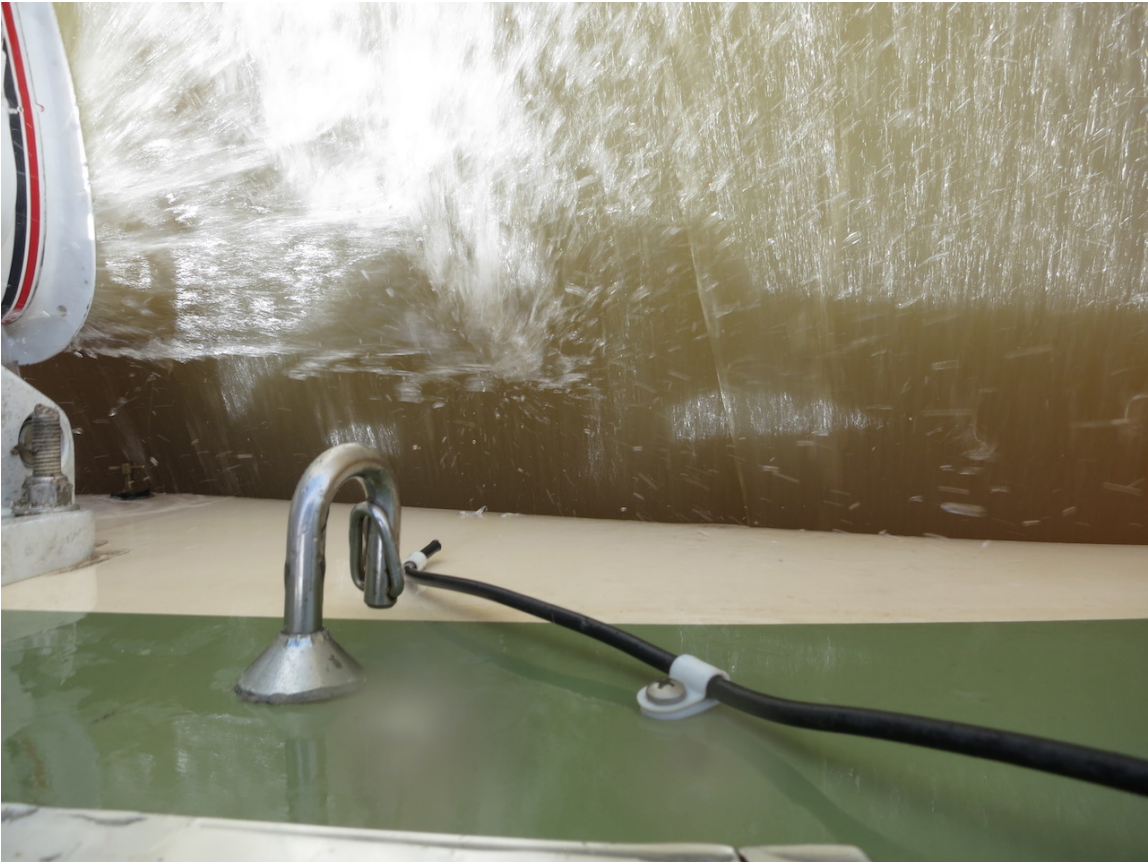

*Fig. 4* was taken at U=8 mph and shows the transom completely dry with the water separating as a sheet parallel to the boat bottom at the transom. This is the onset of lift (Kutta condition).

First, the lift force is not necessarily quadratic in the boat speed U because the lift coefficient may depend on other dimensionless parameters that depend on U. The most important of those in our case is the Froude nr. $F = U/\sqrt{gd}$, where d is submerged depth measured along the side

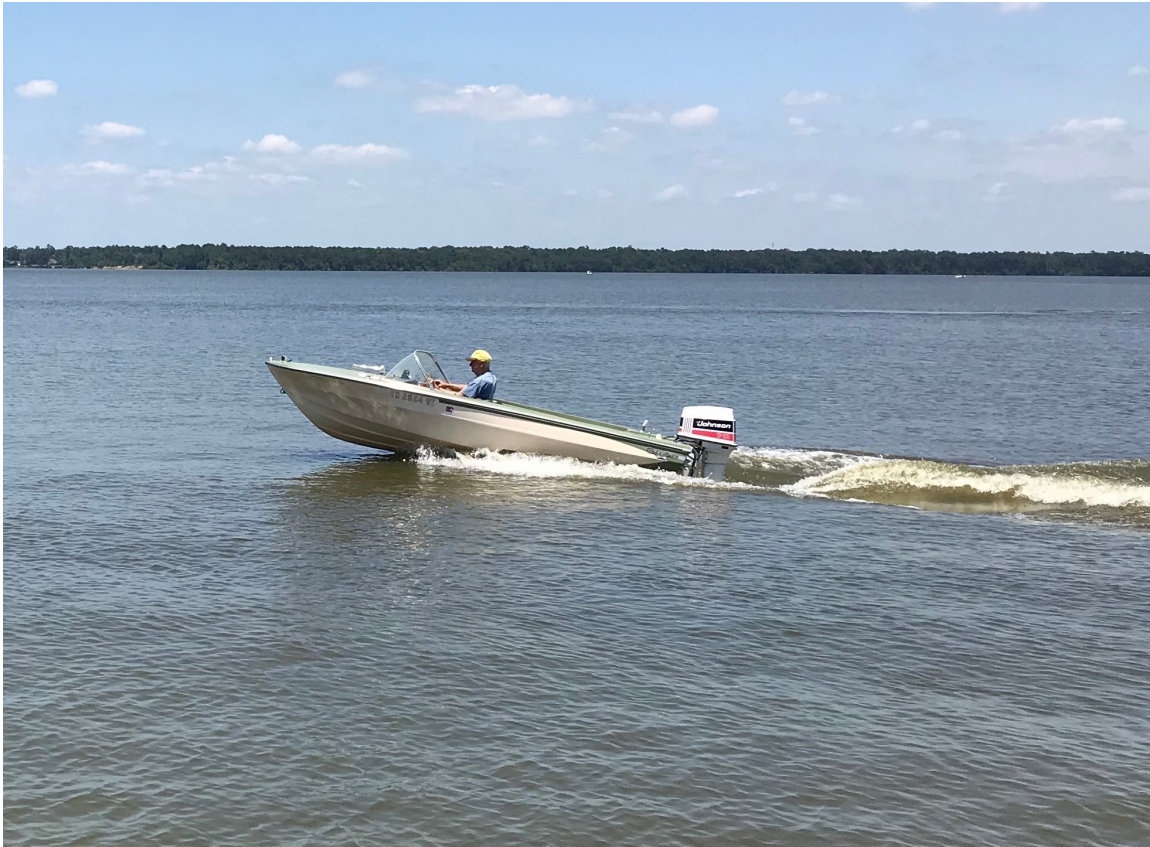

*Fig. 5* shows the same boat plowing with large trim angle α = 5.8° at the onset of lift, U≈8 mph, d≈lα≈1' and F≈2.3 (photo by Cornelia Küffner).

at the transom. Roughly seen, the Froude nr. measures the ratio of the kinetic to the gravitational potential energy. We can measure the submerged depth d≈lα at the transom by estimating the wet length l and while measuring the boat's trim angle α. The wet length l can be estimated from photographs. The boat's trim angle α can be measured as a function of speed by placing a hand-held digital level on the deck of the boat while knowing the difference in angle between deck and running surface. Unlike the

Reynolds nr. the Froude nr. is not universal, it depends on the details of boat bottom shape.

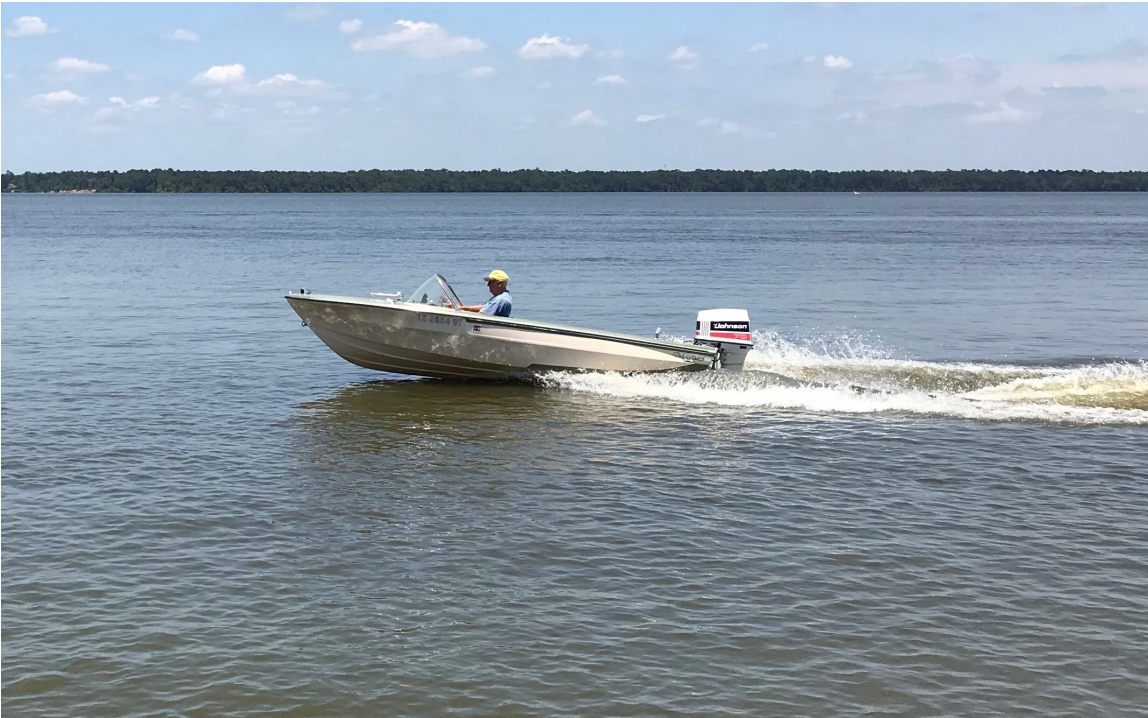

*Fig. 6* shows the boat at planing speed U≈14 mph, $\alpha = 6.7°$, d≈l$\alpha$≈.9', and F≈4 (photo by Cornelia Küffner).

## 4. The Froude number, planing, and the trim angle

Our observations were made using a classic 1968 Glastron v153 boat powered by a 1991 Johnson 70 hp outboard, a digital level to measure angles, and a hand held waterproof GPS to read the speed. At low speed with zero trim angle the water level is the same around the entire boat and buoyancy carries all of the weight. As the boat slowly accelerates then at some speed U the trim angle $\alpha$ of the boat increases from zero as the bow of the boat tries to climb up onto the bow wave (the boat is trying to climb the bow wave in fig. 5). At U ≈ 8 mph, α≈5.8 deg., d≈la≈1' and F≈2.3, the boat plows with lift developed because the transom is completely dry (fig. 4 and 5). See also the 2D calculation by in [7] where the transom becomes dry at F≈2.25.

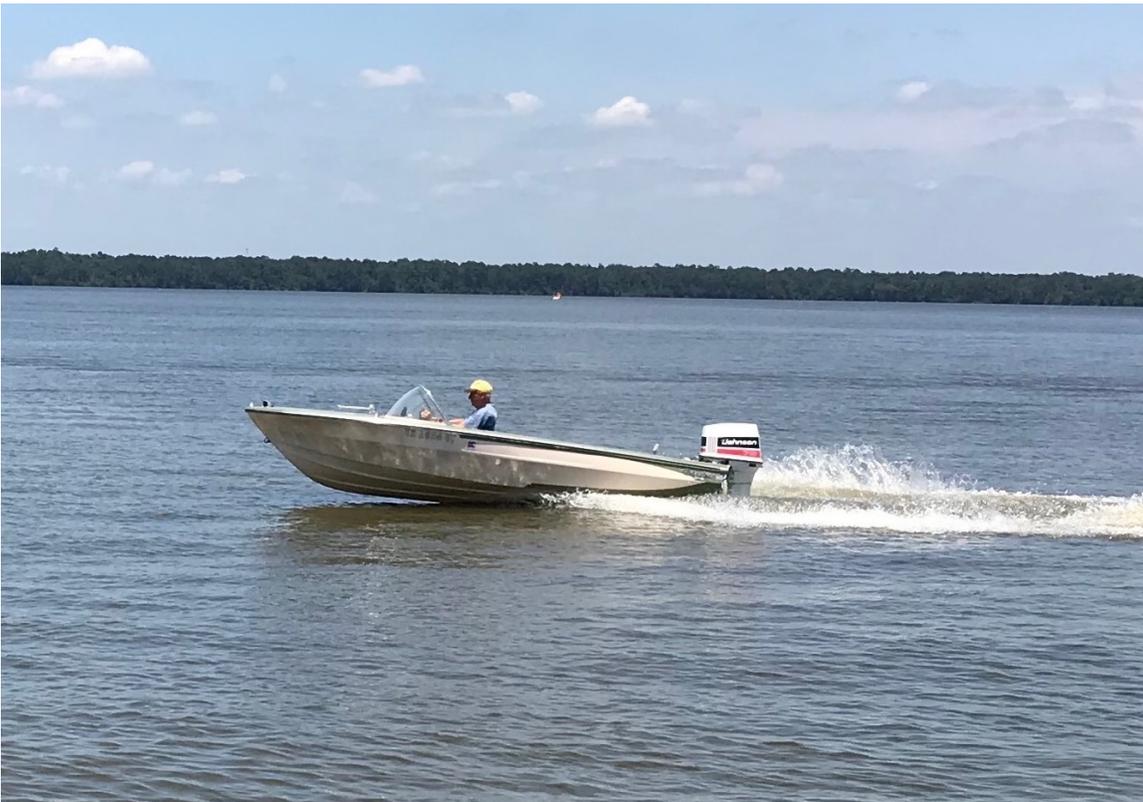

*Fig. 7* shows the boat at 'clean planing speed' U≈18 mph, α = 2°, d≈.3' and F≈9. Here, buoyancy is negligible. Note visually that the wetted depth d is less than in fig. 6 (photo by Cornelia Küffner).

When the speed is increased further to about 14 mph then the boat planes; it has climbed on top of the water and now glides with α ≈ 6.7°, d≈.9' and F≈4. As the speed is further increased to about 18 mph then the trim angle falls to $\alpha \approx 2°$ and remains constant all the way to wide open throttle. At U=18 mph we have d≈.3' and F≈9. The motor has hydraulic trim so that the trim angle of the motor can be changed relative to the boat. In all of these measurements the trim angle of the motor relative to the transom is fixed so that the propeller shaft is parallel to the centerline of the running surface of the boat.

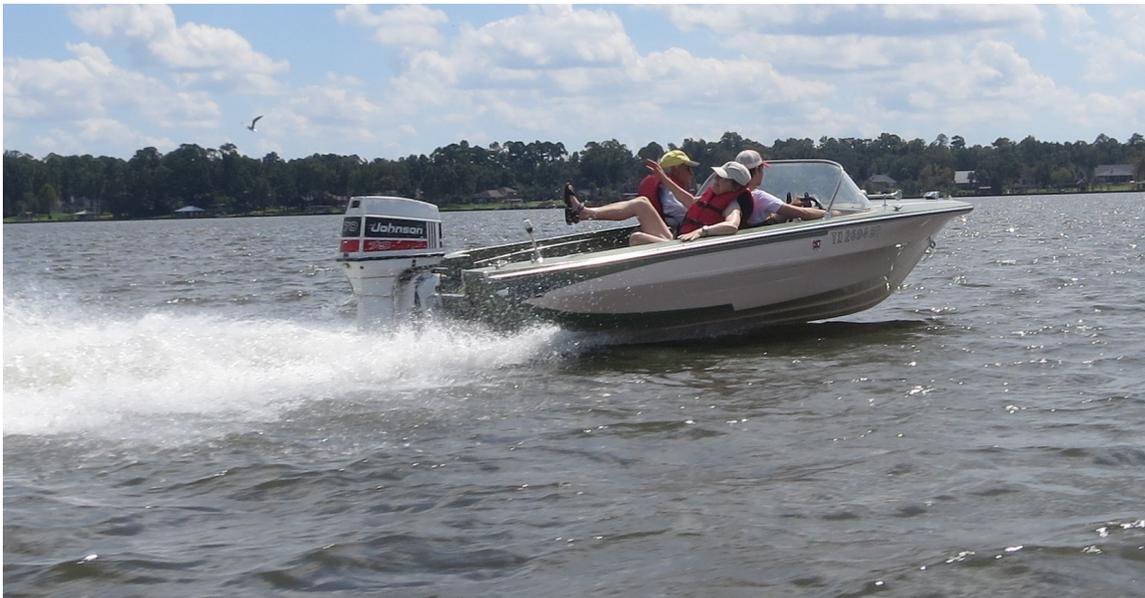

*Fig. 8* shows the v153 planing high on the water at U≈40 mph (photo by Hans Küffner-McCauley).

If we use our result obtained below then the lift coefficient is very nearly $c_L = \alpha$ for a v-bottom boat (prismatic shaped hull) at speeds where the wet running area is small and near the transom, as in fig. 8. At those speeds the wet area is A≈bl/2 while at the low speeds of figures 5-7 one sees from the photos that the wet area is more nearly rectangular with A≈bl.

With buoyant force $F_B = \rho g b l d$ then $F_L / F_B = F^2 c_L / 2$ so lift is large enough to neglect buoyancy when $F >> \sqrt{2/c_L}$. With $\alpha \approx 2°$ we predict that when F>>7.5 buoyancy is negligible. In fig. 8 with l≈4' and d≈.03x4'=.12' we have F≈30. Buoyancy is clearly negligible there.

Gaining speed in the high performance game is largely the search for drag reduction. Air is nearly 800 times less dense than water so the more of the boat (and the motor's gearcase) that you can lift above the water's surface then the faster the boat will go. A tunnel boat (catamaran) is like a wing of very low aspect ratio. There, air lifts the boat in addition to the lift provided by water pressure on the boat's sponsons. Surface-piercing propellers (fig. 2) are required for speeds above about 40 mph. In surface-piercing some fraction of each propeller blade is above the waterline during each rotation, as is the case in figures 2 and 7. The boat shown in fig. 2 (with 75 hp) weighs 840 lb with driver and full rigging and runs about 68 mph on about one square foot of wet surface area.

## 5. The lift coefficient for v-bottom boats and tunnel sponsons

First, we include the effect of deadrise. The transverse cross section of a v-bottom boat has a triangular (prismatic) shape, as is indicated in fig. 2 although in that case the bottom has a flat 'pad' of 7" width along the hull's centerline. The wet surface area of a pure v-bottom (without pad) at high speeds where the boat 'tail rides' is triangular (fig. 7). The deadrise angle β is the angle made by the transverse section of bottom with the horizontal. A flat bottom boat has β=0, the v153 has $β = 18°$. The boat (Allison R14) shown in fig. 2 has $β = 20°$. The deadrise may vary along the boat's centerline, so we mean here the deadrise near the transom, the deadrise along the area of the boat's wet running surface at high speeds. A tunnel boat (catamaran) consists of two sponsons of right triangular (half-prismatic) shape and channels air two dimensionally beneath the deck between the sponsons, providing an air lift coefficient with the highest possible lift, $c_L = 2\pi\alpha$, that of a two dimensional air flow bounded by the two sponson walls. The lift coefficient for one sponson planing on the water is half that of a v-bottom with the same deadrise. Experiments by Clement [8] produced a table of lift coefficients at different trim angles for a v-bottom with $β = 12.5°$. The result is that $c_L \approx \alpha$ for those experiments. Next, we try to model the lift on a v-bottom like the Glastron v153 by the lift on the bottom half of a triangular shaped wing.

The shape of the wet running surface of a v-bottom at high speeds is roughly that of a raked delta wing. The lift is determined by the circulation about the wing, and for a 3-dimensional wing the circulation varies along the span. If one has understood Newman [2] then it's easy to calculate the circulation density Γ(z) along the wingspan, -b/2≤z≤b/2, where the z-axis is along the unraked wing. From that, it's easy to calculate the lift for a delta wing raked upward from the horizontal through an angle β. It's somewhat harder to calculate the induced drag [1,2], but that can also be done to predict the lift to drag ratio $F_L/F_D$, which is the important parameter for a fully submerged hydrofoil of the same shape. A boat bottom, to zeroth order, is like a the bottom half of a hydrofoil of the same shape. We will see that piercing the air-water interface reduces the lift on a hydrofoil (it increases the drag as well). The lift coefficient for half of a raked delta wing [4] is

$$c_L = \pi \alpha \cos\beta / 2. \qquad (4)$$

This is about 1.5 times too large compared with Clement's data [8]. The reason for the discrepancy is that the boat bottom creates lateral spray and waves at the interface, which a fully submerged hydrofoil does not. Any lateral motion of the fluid reduces the lift and increases the drag. Clement's data can be pretty well accounted for by writing

$$c_L \approx \alpha \cos\beta, \qquad (5)$$

and even this can be improved superficially by including a linear factor depending on aspect ratio [4]. The result can be used to understand how the combined water lift on the sponsons and airlift on the deck and tunnel floor account for carrying the total weight of a tunnel boat moving at typical speeds 70-140 mph (depending on power), where F≥≥10 [4].

The analogy of a planing hull bottom with a lifting surface is not new. Newman [2] wrote that one can see the shedding of the trailing/starting vortex by moving a flat plate in the kitchen sink. I was unable to see that so I looked over the transom of our v153 while my younger son drove and recorded the GPS speed. Newman also wrote that planing has some of the features of the lift on the bottom half of a wing but with the complication of running at the air-water interface. Faltinsen [3], following Newman, showed in a model calculation how the lift vanishes if the boat's beam b at the stern vanishes.

## 6. Earlier work on lift and planing

I will comment on some differences with previous discussions of planing as lift that some naval architects have relied on (Savitsky [9]). In Savitsky's paper the Kutta condition is not mentioned so the onset of lift and the reason for the onset are not discussed. Savitsky assumes a lift coefficient $c_L = A\alpha + B\alpha^2$ where he asserts that the first term is for lifting surfaces of high aspect ratio and the second one is assumed (without proof or experimental evidence) to reflect very small aspect ratios with large transverse flow. The formula obeys the condition that the lift vanishes when the trim angle vanishes, but disagrees with Newman, who finds the lift proportional to $\alpha$ even for slender bodies [2]. A body is either of high or low aspect ratio so the formula makes no sense. Savitsky further claims without proof that his formula is approximated by $c_L = C \text{Æ}^{-1/2} \alpha^{1.1}$ where C, like A and B, is a fudge factor. This is impossible: a polynomial cannot be approximated by a function with a branch point and vice versa. Furthermore, there is no known hydrodynamic basis for any exponent other than unity. Savitsky's formulae for drag ignores lateral wave making and spray, which are the main causes of drag.

There are two other ideas of planing as lift that float around in naval architecture literature. Savitsky considers two definitions of planing. First, where the fluid separates from the transom and chines (Kutta condition for onset of lift). We know (see fig. 4) that the eddy-free separation of the flow at the transom is a necessary but insufficient for planing. This defines lift but not planing. Our definition of planing speed (fig. 6) is the slowest speed before a further throttle reduction allows the boat to sink into the water and plow. Clean planing occurs only if there is enough power to make the trim angle and wetted beam smaller as speed is increased. In his fig. 18 he also shows curves (without data points or error bars) of porpoising limits for v-bottoms of different deadrise, but no statement of straightness or firmness of the boats' bottoms is mentioned. In fact, no information is given about the data's origins. Contrary to his graphs you can make any boat porpoise by rockering the bottom. Some boats may porpoise a bit at lower speeds and then level out at high speeds. The graphs of his fig. 18 are unreliable as a basis for predictions.

A 1950s era discussion of planing as lift (Du Cane [10]) is based on a back of the envelope calculation that predicts a lift coefficient proportional to cosα, with finite lift at zero attack angle. This is wrong. Lift must vanish as α goes to zero and increase as α increases, at small angles. The calculation is based purely on leading edge spray for a boat of infinite beam (two dimensional flow) neglecting the side spray and waves that are the main cause of lift-reduction and drag. There is no trailing edge in the calculation, and we know that lift cannot be calculated from a leading edge condition. I.e., there is no Kutta condition in the calculation. The delta wing lift coefficient (4) is based on a three dimensional flow where the Kutta condition was imposed.

## 7. Scaling laws for propellers and speed

Sometimes scaling laws can be obtained by using dimensional analysis and defining classes, as I have done when I derived three very useful scaling laws elsewhere [4,11]. There is a power-speed-weight scaling formula, a scaling relation that predicts propeller diameter based on shaft horsepower and shaft RPM, and a third that to a more limited degree predicts the effect on boat speed of lowering the gear ratio. In all of these cases the boats and gearcases are divided systematically into classes based on drag coefficients, a dimensionless quantity analogous to the lift coefficient. I've used these scaling laws to predict known propeller diameters and existing APBA OPC speed records.

**Acknowledgement**

I'm grateful to Edgar Rose, former head of engineering at OMC, for encouraging me to work on propeller theory after the OPC national championship boat races in 1977. That motivated me to work through Newman's and Landau-Lifshitz's treatments of lift and drag on wings in 1980. A full set of mathematical equations for designing a surface piercing propeller is provided in [4], and has been used with CAD to 'print out' a three dimensional model of a cleaver.